\documentstyle[aaspp4,psfig]{article}

\begin{document}

\def\gsim{\lower0.5ex\hbox{$\; \buildrel > \over \sim \;$}}
\def\lsim{\lower0.5ex\hbox{$\; \buildrel < \over \sim \;$}}
\def\rs{\mbox{$R_{\rm s}$}}
\def\rms{\mbox{$R_{\rm ms}$}}
\def\rh{\mbox{$R_{\rm h}$}}
\def\ro{\mbox{$R_0$}}
\def\ak{\mbox{$a_{\rm k}$}}
\def\nuu{\mbox{$\nu_{\rm u}$}}
\def\nuk{\mbox{$\nu_{\rm k}$}}
\def\num{\mbox{$\nu_{\rm m}$}}
\def\nub{\mbox{$\nu_{\rm b}$}}
\def\dnu{\mbox{$\Delta\nu$}}
\def\ms{\mbox{$M_{\odot}$}}

\title{On the nature of the compact star in 4U 1728-34}

\author{
Xiang-Dong Li\altaffilmark{1},
Subharthi Ray\altaffilmark{2,3,4},
Jishnu Dey\altaffilmark{3,4},
Mira Dey\altaffilmark{2,3}, and
Ignazio Bombaci\altaffilmark{5}
}

\altaffiltext{1}{
Department of Astronomy, Nanjing University, Nanjing 210093, China;
email: lixd@nju.edu.cn
}
\altaffiltext{2}{
Department of Physics, Presidency College, Calcutta 700 073, India;
e-mail : deyjm@giascl01.vsnl.net.in
}
\altaffiltext{3}{
Abdus Salam ICTP, Trieste, Italy
}
\altaffiltext{4}{
Azad Physics Centre, Maulana Azad College, Calcutta 700013, India;
email: azad@vsnl.com
}
\altaffiltext{5}{
Dipartimento di Fisica, Universit\'a di Pisa, via Buonarroti 2, I-56127
and INFN Sezione Pisa, Italy
}

\begin{abstract}

The discovery of kilohertz quasi-periodic oscillations (kHz QPOs) in
low-mass X-ray binaries (LMXBs) with the {\em Rossi X-ray Timing
Explorer} has stimulated extensive studies of these sources. Recently,
Osherovich \& Titarchuk suggest a new model for kHz QPOs and the related
correlations between kHz QPOs and low frequency features in LMXBs. Here
we use their results to study the mass - radius relation for the atoll
source 4U 1728-34. We find that, if this model is correct, 4U 1728-34 is
possibly a strange star rather a neutron star.  

\end{abstract}

\keywords{accretion, accretion disks - stars: neutron 
          - stars: individual (4U 1728-34) - X-ray: stars}

\section{Introduction}

With the {\em Rossi X-ray Timing Explorer} kHz QPOs have been discovered
in about 20 neutron star LMXBs (see van der Klis \cite{vk98} for a
review). In many 
cases, two simultaneous kHz peaks ($\sim 200-1200$ Hz) are observed in
the power spectra of the X-ray count rate variations, with the
separation frequency $\dnu$ roughly constant (e.g., Strohmayer et al.
\cite{s96}; Ford et al. \cite{f97}; Wijnands et al. \cite{w97}).
Sometimes a third kHz peak is detected in a few atoll sources during
type I X-ray bursts at a frequency $\nub$ equal to the separation
frequency $\dnu$ of the two peaks or twice that (see Strohmayer, Swank,
\& Zhang \cite{ssz98} for a review). These two results suggest that a
beat-frequency mechanism is at work, with the upper kHz peak at the
Keplerian orbital frequency at the inner edge of the accretion disk
around the neutron star, the third peak at the neutron star spin
frequency (or twice that), and the lower kHz peak at the difference
frequency between them (Strohmayer et al. \cite{s96}; Miller, Lamb, \&
Psaltis \cite{mlp98}).

Within the beat-frequency model, the burst oscillations $\nub$ and the
kHz peak separations $\dnu$ are both believed to be close to the neutron
star spin frequency and thus $\dnu$ should remain constant. However,
recent observations have challenged this interpretation: in Sco X-1 (van
der Klis et al. \cite{vk97}),
4U 1608-52 (M\'endez et al. \cite{m98}), and 4U 1735-44 (Ford et al.
\cite{f98}), $\dnu$ slowly decreases as the frequencies of both QPOs
increase. Moreover, in 4U 1636-53 the frequeny $\nub$ of the burst
oscillations (or half this value) does not match the frequency
difference $\dnu$ between the kHz QPOs (M\'endez, van der Klis, \& van
Paradijs \cite{mkp98}). In the atoll source 4U 1728-34, M\'endez \& van
der Klis (\cite{mk99}) find that $\dnu$ is always significantly smaller
than $\nub$, even at the lowest inferred mass accretion rate, when
$\dnu$ seems to reach its maximum value, and that $\dnu$ decreases
significantly as the frequency of the lower kHz QPO increases.

A different approach was recently invoked by Osherovich \& Titarchuk
(\cite{ot99a}) and Titarchuk \& Osherovich (\cite{to99}),
who proposed a unified classification of kHz QPOs and the related low
frequency phenomena. In this model, kHz QPOs are modeled as
Keplerian oscillations under the influence of the Coriolis force in a
rotating frame of reference (magnetosphere). The frequencies
$\nuu$ of the upper kHz QPO branch hold a hybrid frequency relation with
the Keplerian frequencies $\nuk$ (referred to the lower kHz QPO branch):
$\nuu^2=\nuk^2+(2\num)^2$, where $\num$ is the rotational frequency of
the star's magnetosphere. For Sco X-1, the QPOs with frequencies
$\sim 45$ and 90 Hz are interpreted as the 1st and 2nd harmonics of the
lower branch of the Keplerian oscillations (Osherovich \& Titarchuk
\cite{ot99a}). The same interpretation is applied to the $\sim 35$ Hz
QPOs in the atoll source 4U 1702-42 (Osherovich \& Titarchuk
\cite{ot99b}). The observed low Lorentzian frequency in 4U 1728-34 is
suggested to be associated with
radial oscillations in the boundary layer of the disk, whereas the
observed break frequency is determined by the characteristic diffusion
time of the inward motion of the matter in the accretion flow (Titarchuk
\& Osherovich \cite{to99}). Predictions of this model regarding
relations between the QPO frequencies mentioned above compare favorably
with recent observations of Sco X-1, 4U 1608-52, 4U 1702-429, and 4U
1728-34.

The discovery of kHz QPO features allows to probe not only the
interaction between the accretion disk and the stellar magnetosphere,
but also the structure of the compact stars involved. In this {\em
Letter}, based on the work of Osherovich \& Titarchuk (\cite{ot99a}) and
Titarchuk \& Osherovich (\cite{to99}), we
study the mass - radius ($M-R$) relation of the atoll X-ray source 4U
1728-34, and show that it is possibly a strange star.

\section{The mass - radius relation for 4U 1728-34}

Titarchuk \& Osherovich (\cite{to99}) suggest a model for the radial
oscillations and diffusion in the viscous boundary layer of the
accretion disk in LMXBs. Their dimensional analysis
has identified the corresponding frequencies which are consistent with
the low Lorentzian and break frequencies for 4U 1728-34 observed by Ford
\& van der Klis (\cite{fk98}), the predicted values for the break
frequency related to the diffusion in the boundary layer are also
consistent with the observed ones for the same source. The presence of
the break frequency and the correlated Lorentzian frequency suggests the
introduction of a new scale in the phenomenon. One attractive feature of
the model is the introduction of such a scale in the model through the
Reynolds number. The best fit for the observed data was obtained when
\begin{equation}
\ak=(M/\ms)(\ro/3\rs)^{3/2}(\nu/364\,{\rm Hz})=1.03,
\end{equation}
where $M$ is the stellar mass, $\ro$ is the inner edge of the accretion
disk, $\rs=2GM/c^2$ is the Schwarzschild radius, and $\nu$ is the spin
frequency of the star. Given the 364 Hz spin frequency of 4U 1728-34
(Strohmayer et al. \cite{s96}),
from Eq. (1) we derive the inner disk radius
\begin{equation}
\ro\simeq 9(\ak/1.03)^{2/3}m^{1/3}\,{\rm km},
\end{equation}
where $m=M/\ms$. Since the innermost radius of the disk must be larger
than the radius of the star itself, this leads to a mass-dependent upper
bound on the stellar radius,
\begin{equation}
R\lsim 9(\ak/1.03)^{2/3}m^{1/3}\,{\rm km},
\end{equation}
which is plotted in dot-dashed curve in Fig.~\ref{fig1}. The dotted
curve in Fig.~\ref{fig1} represents the Schwarzschild radius, which
presents the lower limit of $R$.

Figure \ref{fig1} compares Eq. (3) with the theoretical $M-R$
relations (solid
curves) for nonrotating neutron stars given by six recent realistic
models for the equation of state (EOS) of dense matter. In models
UU (Wiringa, Fiks, \&  Fabrocini \cite{wff88}), BBB1 and BBB2 (Baldo,
Bombaci, \& Burgio \cite{bbb97}), the neutron star core is
assumed to be composed by an uncharged mixture of neutrons, protons,
electrons and muons in equilibrium with respect to the weak interaction
($\beta$-stable nuclear matter). These models are based on microscopic
calculations of asymmetric nuclear matter by use of realistic nuclear
forces which fit experimental nucleon-nucleon scattering data, and
deuteron properties. In model Hyp (Prakash et al. \cite{p97}), hyperons
are considered in
addition to nucleons as hadronic constituents of the neutron star core.
The curve labeled BPAL12 is a very soft EOS for $\beta$-stable
nuclear matter (Prakash et al. \cite{p97}), giving a lower limiting mass
and a smaller
radius with respect to a {\it stiff} EOS. Finally, we
consider the possibility that neutron stars may possess a core with a
Bose - Einstein condensate of negative kaons (e.g. Glendenning \&
Schaffner-Bielich \cite{gs99}).
The main physical effect of the onset of K$^-$ condensation is a
softening of the EOS with a consequent lowering of the neutron star
maximum mass and possibly of the radius. This is shown with the curve
labeled K$^-$. It is clearly seen that only
the upper part of EOSs UU, BBB1, BBB2, and BPAL12 are compatible
with Eq. (3).

Another constraint on the mass and radius of 4U 1728-34 results from the
requirement that the inner radius $\ro$ of the disk must be larger than
the last stable circular orbit $\rms$ around the star, which is given by
(Bardeen, Press, \& Teukolsky \cite{bpt72}),
\begin{equation}
\rms=\rh(3+Z_2-[(3-Z_1)(3+Z_1+2Z_2)]^{1/2}),
\end{equation}
where 
\begin{equation}
Z_1=1+[1-(a/\rh)^2]^{1/3}[(1+a/\rh)^{1/3}+(1-a/\rh)^{1/3}],
\end{equation}
\begin{equation}
Z_2=[3(a/\rh)^2+Z_1^2]^{1/2},
\end{equation}
with $\rh=GM/c^2$ and $a=I\Omega/Mc$
(where $I$ is the moment of inertia and $\Omega=2\pi\nu$ is the angular
velocity of the star, respectively). For a given mass $M$, from Eq. (2) we 
can find $R_0$. From the inequality $\ro\gsim\rms$ and Eqs. (4-6), one can 
calculate the lower limit of $a$, and from there the lower limit of the moment
of inertia $I$.  To account for the inflation of the radius caused by 
rotation, we adopt the formula 
\begin{equation}
I=0.21MR^2(1-\rs/R)^{-1},
\end{equation}
proposed by Burderi et al. (\cite{b99}), to obtain the lower limit on
the stellar radius $R$
\footnote{There are actually two solutions of Eq. (7) with given $I$ and
$M$, one serves as  the upper limit of $R$ and the other the lower limit
of $R$. But the former solution gives an unphysically small radius, and
is rejected.}.
The results are shown in the dashed curve in Fig.~\ref{fig1}. Combining
the
constraints $R\lsim \ro$ and $\ro\gsim \rms$ reveals that the allowed
distribution of the mass and radius of 4U 1728-34 must lie between the
dot-dashed curve and the dashed curve. None of the neutron star EOSs
extends to this region. Thus it seems to rule out the possibility that
4U 1728-34 is a neutron star! Including rotational effects will shift
the theoretical neutron star $M-R$ curves to up-right in
Fig.~\ref{fig1}, making
the contrast between the theoretical neutron star models and the
derived mass and radius range of 4U 1728-34 even worse.

The difficulty of the neutron star model for 4U 1728-34 suggests that it
may belong to another type of compact objects, that is, strange stars,
which are entirely made of deconfined {\it u,d,s} quark matter ({\it
strange matter}).  The possible existence of strange stars is a direct
consequence of the conjecture (Bodmer \cite{b71}; Witten \cite{w84})
that strange matter may be the absolute ground state of strongly
interacting matter.  Detailed studies have shown that the existence of
strange matter is allowable within uncertainties inherent in a strong
interaction calculation (Farhi \& Jaffe \cite{fj84}); thus strange stars
may exist in the universe.

Most of the previous calculations (e.g. Alcock, Farhi, \&  Olinto 
\cite{afo86}; Haensel, Zdunik, \& Schaeffer \cite{hzs86})  of strange
star
properties used an EOS for strange matter based on the phenomenological
nucleonic bag model, in which the basic features of quantum
chromodynamics, such as quark confinement and asymptotic freedom are
postulated from the beginning, though the deconfinement of quarks at
high density is not obvious in the bag model.
Recently, Dey et al. (\cite{d98}) derived an EOS for strange matter,
which has asymptotic freedom built in, shows confinement at zero baryon
density, deconfinement at high density, and gives a stable configuration
for chargeless, $\beta$-stable strange matter. In this model the quark
interaction is described by an interquark vector potential originating
from gluon exchange, and by a density dependent scalar potential which
restores the chiral symmetry at high density. Using the same model (but
different values of the parameters with respect to those employed in Dey
et al. \cite{d98}) we calculated the $M-R$ relations, which are also
shown in solid curves labeled ss1 and  ss2 in Fig.~\ref{fig1},
corresponding to
strange stars with maximum masses of $1.44 \,\ms$ and $1.32\,\ms$ and
radii of 7.07 km and  6.53 km, respectively. It is seen that the region
confined by the dashed and dot-dashed curves for 4U 1728-34 in
Fig.~\ref{fig1} is
in remarkable accord with the strange star models with masses $\lsim 1.1
\ms$. Figure \ref{fig1} clearly demonstrates that a strange star model
is more compatible with 4U 1728-34 than a neutron star one. This seems
to be important since the model of Dey et al. (\cite{d98}) employs quark 
masses which
decrease with increasing density, a manifestation of chiral symmetry 
restoration (CSR), and the parametrization of CSR can be deduced from the 
mass and radius of the strange star fitted. 

The lower limit on the moment of inertia deduced above does not depend on the 
EOS. In particular Eq. (7) is an extrapolation from neutron stars and it may 
be doubted whether it is valid for strange stars. We have checked that the 
standard calculated moment of inertia of the strange star of mass $M \simeq
1.1 M_{\odot}$ agrees with Eq. (7) with an error of about $20\;\%$, and the 
numerical factor 0.21 can change to 0.28 (ss1) or 0.26 (ss2) depending on the 
model parameters. The moment of inertia calculation of the strange star with 
variable density layers may not in fact satisfy the assumptions of the 
standard calculation.  It is satisfying to find that an EOS-independent 
estimate can already reproduce the inertia to a reasonable accuracy.

\section{Discussion and conclusion}

We have shown that strange stars are more consistent with the properties
of 4U 1728-34 compared to neutron stars, {\it if} the model of
Osherovich \& Titarchuk (\cite{ot99a}) and Titarchuk \& Osherovich
(\cite{to99}) correctly interprets the high- and low-frequency QPO
phenomena in 4U 1728-34. Since our analysis does not involve any
unknown physical factors such as the magnetic field strength and
topology, radiation processes, and
the mass accretion rate, it seems to provide the first definite evidence
for the existence of strange stars. It remains to see whether other
compact stars in LMXBs would be identified as strange stars from
observations of their QPO phenomena. If so, there will be very deep
consequences for both physics of strong interactions and astrophysics.

\begin{figure}
\centerline{\psfig{figure=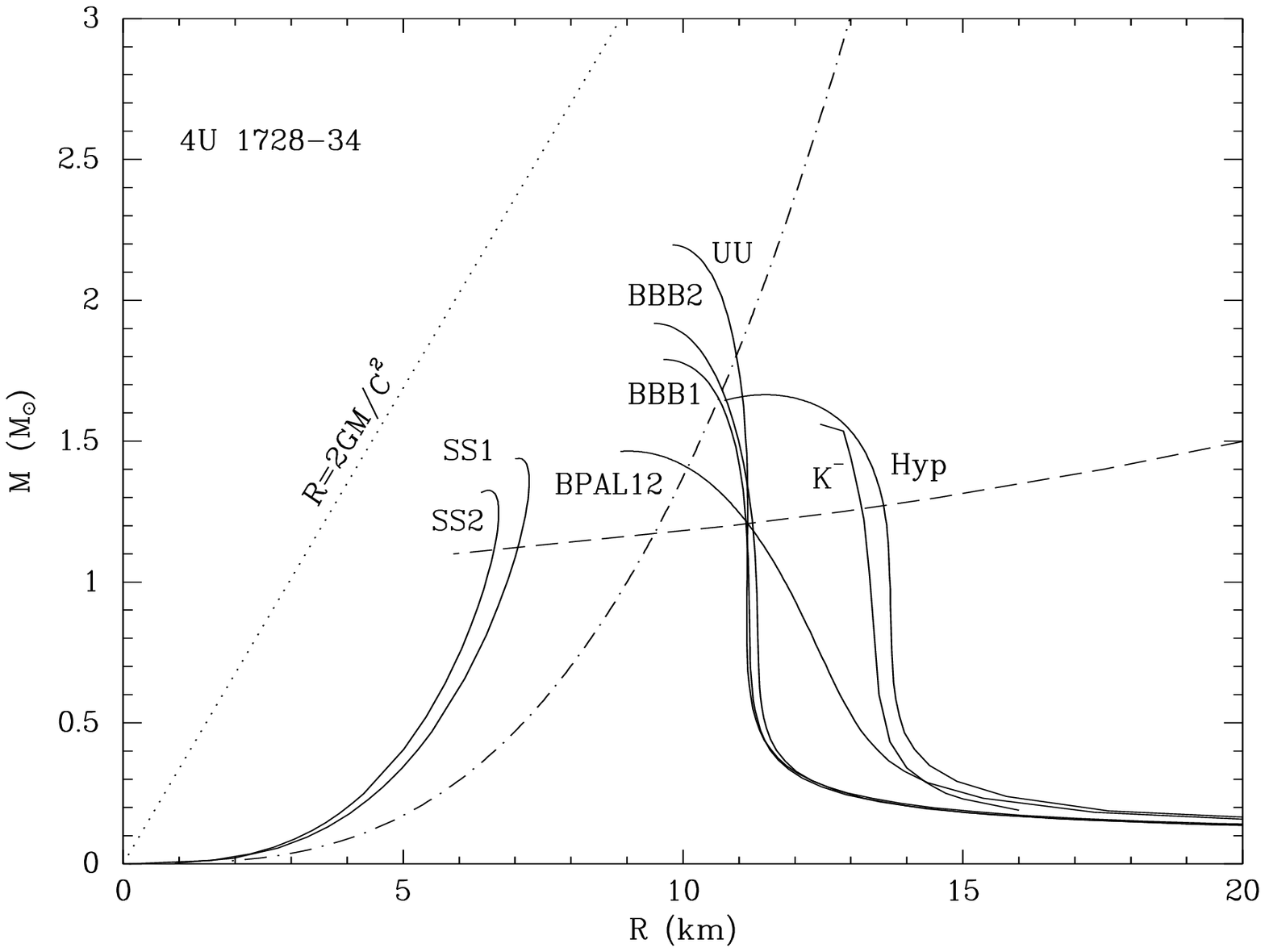,angle=-90}}
\caption{
Comparison of the $M-R$ relation of 4U 1728-34 determined from RXTE
observations with theoretical models of neutron stars and of strange
stars. The range of mass and radius of 4U 1728-34 is allowed in the
region outlined by the dashed and dot-dashed curves. The solid curves
labeled UU, BBB1, BBB2, BPAL12, Hyp, and K$^-$ represent various $M-R$
relations for {\it realistic} EOSs of nonrotating neutron stars; the
solid curves labeled ss1 and ss2 are for strange stars.
}
\label{fig1}
\end{figure}

Of course our conclusion is subject to the uncertainty in the parameter
\ak, which is determined by the best fitting of the observed data.
Increase or decrease in $\ak$ will enlarge or reduce the allowed region
of the mass and radius of 4U 1728-34 in Fig.~\ref{fig1}. 
For example, if $\ak=2$, nearly
all neutron star EOSs are compatible with 4U 1728-34. However, according
to Titarchuk \& Osherovich (\cite{to99}), the dependence of $\chi^2$ on
$\ak$ is quite strong:
$\chi^2=38024-73076\ak+35732\ak^2\simeq 35732[(\ak-1.03)^2+0.019]$. A
$\sim 15\%$ variation of $\ak$ from the preferred value 1.03 will double
the value of $\chi^2$, greatly reducing the confidence level of the fit.
Thus we expect that the possible change in $\ak$ and in the derived $M-R$
relation is rather small.

If 4U 1728-34 is a strange star, Fig.~\ref{fig1} reveals that it
possesses a mass $\lsim 1.1\ms$, which is quite small compared to the
$\sim 1.4\ms$ masses of standard neutron stars like PSR 1913+16, though
still barely
consistent with the estimated masses of quite a few radio pulsars within
$68\%$ confidence limit (see Thorsett \& Chakrabarty \cite{tc99}). Since
the mass accretion rates in LMXB atoll sources are lower than the
Eddington limit accretion rate by $1-2$ orders of magnitude, it is
conventionally thought that in such systems almost all
of the matter transferred from the companion stars should be accreted by
the compact stars. The light mass of 4U 1728-34, however, seems to
support the conclusion of Thorsett \& Chakrabarty (\cite{tc99}) that
there is no evidence for extensive mass accretion occurred in LMXBs,
implying that even the simplest evolution of LMXBs is far from being
fully understood. However, formation of strange stars may be
different from that of neutron stars (i.e. type II/Ib supernova
explosions) concerning the glitching phenomena observed in some radio
pulsars (e.g. Alpar \cite{a87}), suggesting that some exotic formation
channels may be at work. Accreting massive neutron stars in binary
systems are not favorable progenitor candidates of strange stars, at
least in the case of 4U 1728-34, since it requires too large mass
($\gsim 0.7\ms$) ejected during the conversion as the neutron star
masses exceed $\gsim 1.8\ms$. Accretion-induced collapse of white dwarfs
remains an open possibility of strange star formation, and could be
responsible for the light mass of 4U 1728-34. The detailed investigation
of this subject is beyond the scope of this paper and wil be discussed
elsewhere.

\acknowledgements
We are grateful to Lev Titarchuk for helpful and prompt e-mails.
X. L. was supported by National Natural Science Foundation of China.
S. R., M. D. and J. D. were supported in part by DST grant no.
SP/S2/K18/96,
Govt. of India.

\clearpage

\end{document}